\documentclass[conference]{IEEEtran}

\usepackage{amsmath}
\usepackage{amssymb}
\usepackage{amsthm}
\usepackage{mathtools}
\usepackage{breqn}

\usepackage[utf8]{inputenc}
\usepackage[T1]{fontenc}
\usepackage{textcomp}
\usepackage{gensymb}
\usepackage{upquote}

\usepackage{graphicx}
\usepackage{xcolor}
\usepackage{soul}

\usepackage{tabulary}
\usepackage{booktabs}

\usepackage{algorithm}
\usepackage{algorithmic}

\usepackage{url}
\usepackage{cite}
\usepackage{pifont}

\usepackage[top=1in, bottom=1in, left=0.7in, right=0.7in]{geometry}

\newcommand{\cmark}{\ding{51}}
\newcommand{\xmark}{\ding{55}}

\begin{document}

\title{RL-based Network Slice Embedding over Space Division Multiplexed Elastic Optical Networks}

\author{\IEEEauthorblockN{Divya Khanure \textsuperscript{$*$}, Riti Gour\textsuperscript{$\dagger$}, Congzhou Li\textsuperscript{$*$}, and Jason P. Jue\textsuperscript{$*$} }

\IEEEauthorblockA{\textsuperscript{$*$}Department of Computer Science\\
The University of Texas at Dallas, Richardson, Texas 75080, USA\\
\textsuperscript{$\dagger$} Department of Aviation and Technology,\\ San Jose State University, San Jose, California 95192, USA
}\\
\vspace{-0.5 cm}
Email:\{divya.khanure, congzhou.li, jjue\}@utdallas.edu, riti.gour@sjsu.edu}

\maketitle

\begin{abstract}
Network slicing over space-division-multiplexed elastic optical networks (SDM-EONs) requires jointly managing spectrum, spatial cores, and compute resources, a coupling that many existing studies ignore by treating compute placement independently from routing and spectrum decisions. This disconnect can cause the spectrum to be allocated along a path, only for the request to fail due to insufficient compute resources along the path, or may result in compute resources being allocated without consideration for spectrum resource availability on the path between compute nodes. We propose a path-constrained reinforcement learning framework that addresses compute node selection and RMCSA, being aware of both resources, restricting the RL agent's action space to nodes along $k$-shortest paths between request endpoints. Training incorporates reward shaping to improve robustness under high load. We propose PPO-Full (Proximal Policy Optimization-Full), which jointly selects compute nodes and routing paths via a multi-dimensional action space, against distance-based heuristics, a greedy baseline, and a decoupled VONE-DRL baseline on a 24-node USNET topology under hotspot traffic conditions. Results demonstrate consistent improvements in acceptance rate over all baselines at high load, with gains becoming more pronounced as traffic intensity increases.

\end{abstract}

\begin{IEEEkeywords}
Network Slicing, Slice Embedding, RMCSA, Reinforcement Learning, Elastic Optical Networks, Hotspot Traffic, Multi-dimensional Action Space.
\end{IEEEkeywords}

\IEEEpeerreviewmaketitle

Supporting high-bandwidth heterogeneous services over a shared optical infrastructure has driven growing interest in network slicing over space-division-multiplexed elastic optical networks (SDM-EONs), where slice embedding demands simultaneous management of spectrum, spatial cores, and compute resources, a joint dependency that many RMCSA formulations overlook~\cite{2431731}. In SDM-EONs, spectrum assignment must satisfy contiguity and continuity requirements, core selection must account for inter-core crosstalk, and compute placement must respect node capacity limitations, making joint optimization computationally intractable. Prior work has largely decoupled compute placement from routing and spectrum assignment, producing suboptimal 
mappings under dynamic conditions~\cite{9309337, 6679238, 10288372}. Furthermore, slice embedding studies enforce strict one-to-one virtual-to-substrate node mappings that limit compute utilization. Coordinated and
co-location-aware strategies in broader VNE research have shown improved acceptance ratios~\cite{NodeFusion, ChowdhuryCoordinatedVNE}, but such principles have not been widely incorporated into SDM-EON slicing frameworks.

Building on our prior heuristic-based work~\cite{ICCCN2026}, which proposed Direct Path-Driven Slice Mapping (DPSM) and Waypoint-Assisted Multi-Segment Slice Mapping (WMSM), this paper incorporates reinforcement learning into the resource allocation process. DPSM assigns compute to the first feasible node along the path; its extended version targets a node near the path midpoint to balance segment lengths. WMSM decomposes requests into shorter independently provisioned segments via compute-capable waypoints; WMSM-F extends this by jointly minimizing path length and segment imbalance. While these heuristics perform well, their deterministic rules cannot adapt to dynamic traffic, motivating the present work, which retains WMSM-F as a greedy baseline and replaces fixed allocation logic with a learned RL policy.

We propose PPO-Full, a Proximal Policy Optimization (PPO) based approach for slice embedding in SDM-EONs that jointly optimizes compute node selection and RMCSA by 
restricting the agent's action space to nodes along $k$-shortest paths, 
making allocation decisions aware of both compute and spectrum 
resources simultaneously. PPO-Full is evaluated against distance-based 
heuristics, WMSM-F, and a decoupled VONE-DRL 
baseline~\cite{DohertVONE2023} on a 24-node USNET topology under 
hotspot traffic conditions. The main contributions are:
\begin{itemize}
    \item A PPO-based slice embedding approach (PPO-Full) that jointly 
    optimizes compute node and routing path selection via a multi-dimensional action space \texttt{MultiDiscrete([$C$, $K$])}, where $C$ denotes the number of candidate compute nodes and $K$ the number of candidate routing paths. This couple computes placement with RMCSA within a single learned policy, enabling the agent to simultaneously reason about compute availability and spectrum efficiency. In our implementation, $C =  K = 3$, giving \texttt{MultiDiscrete([3,3])}. A node-only reranking variant (PPO-Rank) with path fixed to $k=3$ was also explored as an ablation to quantify the value of path selection freedom, confirming that joint node-path selection provides measurable benefit.

    \item A candidate-aware observation space that encodes per-candidate 
    compute availability and segment spectrum state, enabling the agent 
    to make informed joint compute-spectrum placement decisions. The agent selects from $k = 3$ pre-computed shortest paths, chosen empirically after sweeping $k \in \{2, 3, 4, 5\}$, and a variable set of candidate compute nodes filtered by compute feasibility and ranked by path distance, with the top-$N_c$ candidates considered per request.

    \item A consecutive co-location scheme for VNF placement, where 
    multiple consecutive VNF demands $\{f_i, \ldots, f_j\}$ may be 
    co-located on a single physical compute node, and all valid 
    consecutive partitions of $F$ VNFs across 1--$F$ nodes are attempted 
    exhaustively. A quantitative analysis of co-location impact is deferred to an 
extended journal version due to page constraints.
\end{itemize}

The remainder of this paper is organized as follows. Section~II reviews 
related work. Section~III presents the system model. Section~IV 
describes the proposed framework. Section~V presents numerical results. 
Section~VI concludes the paper.

\section{Background Literature}

Early VNE formulations by Gong and Zhu~\cite{6679238} and Wang and 
Hu~\cite{9309337} improved spectral efficiency but were limited to 
single-core architectures without compute-aware placement. Multi-core 
fiber introduced RMCSA research with IC-XT constraints: Zhang~\textit{et 
al.}~\cite{9874982}, Jin~\textit{et al.}~\cite{10209960}, and 
Tang~\textit{et al.}~\cite{Tang2020} proposed crosstalk-aware SDM-EON 
embedding, while Heera~\textit{et al.}~\cite{heera_crootalk_rmcsa, 
heera_congestion_rmcsa} introduced congestion-aware RMCSA heuristics, that is, all treating compute placement as independent from optical allocation.

RL has emerged as a promising alternative for optical resource 
allocation. Chen~\textit{et al.}~\cite{DeepRMSA} proposed DeepRMSA for 
RMSA in single-core EONs, Pinto-R\'{i}os~\textit{et 
al.}~\cite{PintoRios2023} extended DRL to MCF-EONs, and 
Xu~\textit{et al.}~\cite{Xu2022GCN} combined GCN with deep RL for 
topology-aware RMSA. However, none incorporate compute placement, 
limiting applicability to joint compute-spectrum slice embedding.

On the VNE side, Wang~\textit{et al.}~\cite{NodeFusion} and 
Chowdhury~\textit{et al.}~\cite{ChowdhuryCoordinatedVNE} proposed 
co-location and coordinated node-link mapping strategies. 
Gu~\textit{et al.}~\cite{10288372} applied deep RL to RAN slice 
migration at higher architectural layers, while Doherty~\textit{et 
al.}~\cite{DohertVONE2023} applied masked DRL to VONE over single-core 
EONs with decoupled agents, which we adapt as a baseline. Our prior 
work~\cite{ICCCN2026} addressed heuristic-based SDM-EON slice embedding 
with compute-spectrum coordination, the present work replaces 
deterministic heuristics with a path-constrained RL agent, as 
summarized in Table~\ref{tab:reff}.
\begin{table}[htbp]
\centering
\caption{Comparison of related works by SDM-EON, coordination, 
co-location, and solution type}
\label{tab:reff}
\setlength{\tabcolsep}{2pt}
\renewcommand{\arraystretch}{1.05}
\begin{tabular}{p{2.7cm}ccc p{2.3cm}}
\toprule
\textbf{Reference(s)} &
\textbf{SDM-EON} &
\textbf{Coord.} &
\textbf{Co-loc.} &
\textbf{Solution Type(s)} \\
\midrule
\cite{9831362,10209960,2593479,Tang2020,heera_crootalk_rmcsa,
heera_congestion_rmcsa}
& \cmark & \xmark & \xmark & ILP, Heuristic \\
\cite{ChowdhuryCoordinatedVNE,10288372}
& \xmark & \xmark & \cmark & RL, Heuristic \\
\cite{NodeFusion}
& \xmark & \cmark & \cmark & Heuristic \\
\cite{6679238,9309337}
& \xmark & \xmark & \xmark & ILP, Heuristic \\
\cite{DeepRMSA, PintoRios2023, Xu2022GCN}
& \xmark & \xmark & \xmark & RL \\
\cite{DohertVONE2023}
& \xmark & \xmark & \xmark & RL \\
\cite{ICCCN2026}
& \cmark & \cmark & \cmark & Heuristic \\
\midrule
\textbf{Proposed}
& \cmark & \cmark & \cmark & Heuristic + RL \\
\bottomrule
\end{tabular}
\end{table}

\section{System Model and Problem Formulation}

We model an SDM-EON as a graph $G = (V, E)$, where each link supports $|\Phi_e|$ spatial cores with $f$ spectrum slots each, and a subset $V_c \subseteq V$ of compute-capable nodes with finite capacity. Cores are grouped by spatial adjacency to mitigate inter-core crosstalk.

\subsection{Definitions and Operational Semantics}

\textbf{Compute \& OEO Semantics.} Compute capacity is expressed in normalized units representing VNF execution. Each request places $|F|$ VNFs with independent demands, co-located or distributed across path nodes. Deploying a VNF triggers OEO conversion, terminating the incoming lightpath, and relaxing spectrum continuity to within each segment only, allowing decomposition into shorter independently provisioned parts.

Each request $r = (s, d, b_{req}, F)$, where $F = \{f_1, f_2, ..., f_n\}$ is accepted when VNFs are placed at feasible compute nodes and spectrum is provisioned on all resulting segments, with routing, spectrum, and compute decisions coupled sequentially.

\subsection{Cost Model}

The average provisioning cost per accepted VNR is defined as:
\begin{equation}
C_{r_i} = \frac{\sum_{v \in V_c} c_{v,r_i}\,\mathbb{C}_{compute} + 
\sum_{e \in E} \gamma_{e,r_i}\,\mathbb{C}_{spectrum}}{\text{accepted VNRs}},
\end{equation}
where $c_{v,r_i}$ is the sum of compute units actually allocated across 
all VNFs at node $v$, $\gamma_{e,r_i}$ is the total spectrum slots used 
across all RMCSA path segments for request $r_i$, and 
$\mathbb{C}_{compute} = \mathbb{C}_{spectrum} = 1.0$ (equal weighting). 
This metric captures the average total resources, that is, compute units plus 
spectrum slots, consumed per successfully provisioned VNR. A lower 
value indicates more resource-efficient allocation, a method that accepts 
the same requests but routes them more efficiently will incur lower cost.
The overall objective is to minimize this average cost while maximizing 
the acceptance rate.


\subsection{Constraints}

The slice mapping must satisfy the following constraints, else the request is blocked:

\begin{itemize}
\item \textbf{(C1) Modulation reach:} The total path length  for a segment must satisfy $\sum_{e \in P_{r_i}^k} D_e \leq D_{\max}\left(M_{r_i}^k\right), \quad \forall r_i,\; \forall k$, where $P_{r_i}$ is the selected path, $D_e$ is the distance of link $e$, and $D_{\max}(M_{r_i})$ is the maximum transmission reach for modulation format $M_{r_i}$. Three modulation formats are supported: 16-QAM ($\leq$500~km, 1~slot/Gbps), QPSK ($\leq$1000~km, 2~slots/Gbps), and BPSK ($\leq$2000~km, 4~slots/Gbps).

\item \textbf{(C2) Spectrum contiguity:} Each request $r_i$ occupies a 
contiguous slot set $ \Gamma_{r_i,e,\phi} = \{s_{r_i}, s_{r_i}+1, \dots, s_{r_i}+f_{r_i}-1\}, \quad \forall r_i,\; \forall e \in P_{r_i}^k $, where $f_{r_i} = b_{req} \times \eta(M_{r_i})$ is the number of slots per Gbps for the chosen modulation format.

\item \textbf{(C3) Spectrum continuity:} Identical contiguous slots are assigned on all links along a path: $\Gamma_{r_i,e,\phi} = \Gamma_{r_i,e',\phi}, \quad \forall e,e' \in P_{r_i}.$ Continuity across different segments $k$ is not required due to OEO conversion at compute waypoints.

\item \textbf{(C4) Compute feasibility:}
$ c_{v,r_i}^{(j)} \geq f_j, \forall r_i,\; $ 
\text{such that} 
$ \sum_{r_i} \left(c_{v,r_i}^{(1)} + c_{v,r_i}^{(2)} + \cdots + c_{v,r_i}^{(n)}\right) \leq C^{\text{comp}}_{v}, \quad \forall v \in V_c $, 
where $c_{v,r_i}^{(j)}$ is the compute allocated to VNF $j$ of request $r_i$ at node $v$, and $C^{\text{comp}}_{v}$ is the node capacity.

\item \textbf{(C5) Core-group selection (crosstalk avoidance):} Each request is assigned a core $\phi_{r_i} \in \mathcal{G}_k$, where $\mathcal{G}_k$ denotes a non-adjacent core group (e.g., $\mathcal{G}_1 = \{1,3,5\}$, $\mathcal{G}_2 = \{2,4,6\}$, $\mathcal{G}_3 = \{7\}$), ensuring $\sum_k g_{r_i,k} = 1$, with $g_{r_i,k} = 1$ if $\mathcal{G}_k$ is used by $r_i$ and $0$ otherwise.
\end{itemize}


\subsection{Performance Metrics}

\textbf{Acceptance rate} measures the fraction of successfully
provisioned requests. \textbf{Spectrum utilization} is $U =
\frac{\sum_{i}\sum_{e} \gamma_{e,r_i}}{\mathbb{N}}$, where $\mathbb{N}$
is total network slots (links $\times$ cores $\times$ slots/core).
\textbf{Provisioning cost} is the cumulative cost associated with compute units plus spectrum slots consumed. \textbf{Spectrum fragmentation} uses the
external fragmentation index~\cite{wright2015fragmentation}: $F_{ex}(l,c)
= 1 - B_{max}/F_{total}$, where $B_{max}$ is the largest contiguous free
block and $F_{total}$ is the total free slots, averaged across cores as
$F_{link}(l) = \frac{1}{C}\sum_{c=1}^{C} F_{ex}(l,c)$.

The slice embedding problem is NP-hard, generalizing RSA (reducible to
wavelength assignment~\cite{ChatterjeeTutorial}) and VNE (equivalent to
subgraph isomorphism~\cite{ChowdhurySurveyVNE}), with modulation reach,
inter-core crosstalk, and compute placement, adding multidimensional
bin-packing complexity~\cite{BianzinoSurveyNP}, making exact optimization
intractable and motivating the RL-based approach proposed here. PPO-Full's inference time per request ($\approx$135~$\mu$s for the 
policy forward pass on CPU) is comparable to the more complex 
heuristics such as Equidistant ($\approx$1,115~$\mu$s) and VONE-DRL 
($\approx$1,256~$\mu$s), with the computational cost of training 
incurred offline only.

\section{Proposed Framework}

PPO-Full is proposed for network slice embedding in SDM-EONs, generating 
a small set of candidate compute nodes ranked by path distance and using 
a learned PPO policy to jointly select a compute node and a routing path. 
Restricting the candidate set to $C=3$ nodes and $K=3$ pre-computed 
shortest paths strikes a practical balance between decision quality and 
action space tractability: a larger candidate set exponentially increases 
the number of joint node-path combinations the agent must explore, 
leading to slower convergence and sparser rewards, while $K=3$ retains 
sufficient diversity to cover the dominant trade-offs between compute 
proximity, path length, and spectrum availability. $K$ was selected 
empirically after sweeping $K \in \{2, 3, 4, 5\}$, with $K=3$ offering 
the best trade-off between candidate diversity and action space 
tractability. PPO-Full jointly selects a compute node and a routing path 
via \texttt{MultiDiscrete([C,P])}, yielding $C \times K = 9$ joint 
node-path combinations per decision step. A node-only variant 
(PPO-Rank) with path fixed to $k=3$ was explored as an ablation, 
confirming that joint node-path selection provides measurable benefit 
over node reranking alone (results not included in the paper due to space constraints). VNF demands are partitioned consecutively 
across compute nodes starting from the RL-selected anchor node, with 
remaining VNF groups greedily assigned to the nearest feasible nodes. 
PPO-Full is evaluated against distance-based heuristics, a greedy 
baseline, and a decoupled VONE-DRL baseline described at the end of 
this section.

\vspace{-5 pt}

\subsection{Action Space}

PPO-Full operates over $C = 9$ candidate compute nodes ranked by total 
path distance $distance(s,n) + distance(n,d)$ from the $K=3$ pre-computed shortest paths and filtered by compute feasibility, reducing node selection from $O(|V_c|)$ to $O(C)$. It uses \texttt{MultiDiscrete([3,3])}, jointly selecting a compute node and a routing path, yielding $9$ joint node-path combinations per decision step.


\vspace{-5 pt}

\subsection{Observation Space}

PPO-Full observes a normalized feature vector constructed per arrival 
event, comprising: request features ($s$, $d$, $b_{req}$, $f_1 \ldots 
f_F$), network utilization, and network fragmentation ($2+F+2$ values), 
and per-candidate features, which are, compute availability ratio, normalized 
distances to source and destination, segment utilization, maximum free 
block, and fragmentation for each of the two resulting segments (9 
values per candidate). For $C=3$ compute node candidates, $K=3$ path 
candidates, and $F=4$ VNFs, this yields a $(2+F+2) + C \times 9 = 35$-dimensional observation vector.

\subsection{Reward Shaping}

The reward function incentivizes acceptance while penalizing 
spectrum-inefficient allocations:
\begin{equation}
\mathcal{R} = \begin{cases} 
5.0 + 2.0 \times \left(1 - \frac{\min(S, 200)}{200}\right) & \text{accepted} \\
-5.0 & \text{blocked}
\end{cases}
\end{equation}
where $S$ is the total number of spectrum slots consumed by the 
accepted request. The acceptance reward is augmented by a slot-efficiency 
bonus that decreases linearly with spectrum usage, reaching zero for 
allocations consuming 200 or more slots. This encourages the agent to 
prefer compact, spectrum-efficient placements without explicitly 
penalizing fragmentation, simplifying the reward signal compared to 
multi-term formulations while retaining the incentive for resource-efficient 
provisioning.

\subsection{PPO Agent and Hyperparameters}

PPO-Full is implemented using PPO~\cite{10364638} via 
\texttt{stable-baselines3} with an Multi-Layer Perceptron (MLP) policy with two hidden layers with 256 neurons in each layer (\texttt{net\_arch=[256,256]}), well-suited for discrete action spaces 
under non-stationary traffic. Hyperparameters are summarized in 
Table~\ref{tab:ppo}. The best checkpoint is selected based on mean 
acceptance rate evaluated every 50,000 steps at $\lambda \in \{15, 
30, 50\}$. 

\vspace{-8pt}
\begin{table}[htbp]
\centering
\caption{PPO Hyperparameters}
\label{tab:ppo}
\renewcommand{\arraystretch}{1.05}
\setlength{\tabcolsep}{4pt}
\begin{tabular}{lclc}
\toprule
\textbf{Hyperparameter} & \textbf{Value} & 
\textbf{Hyperparameter} & \textbf{Value} \\
\midrule
Learning rate       & $3 \times 10^{-4}$    & Clip range            & 0.2 \\
Batch size          & 256                   & Entropy coefficient   & 0.01 \\
Steps per update    & 2048                  & Training timesteps    & 1,000,000 \\
Epochs per update   & 10                    & Parallel environments & 4 \\
Discount $\gamma$   & 0.99                  & GAE $\lambda$         & 0.95 \\
\bottomrule
\end{tabular}
\end{table}

\vspace{-8 pt}

\begin{algorithm}[H]
\caption{PPO-Full Allocation}
\label{alg:enhanced_rl}
\begin{algorithmic}
\REQUIRE $(s, d, b_{req}, f_1 \ldots f_F)$, $C=3$ candidate nodes 
$\mathcal{C}$, $K=3$ paths $\mathcal{P}_{sd}$, policy $\pi_\theta$
\ENSURE Allocation decision, spectrum and compute assignment
\STATE Build observation $o$ from network state, request features, 
and per-candidate segment features
\STATE Query policy: $a \leftarrow \pi_\theta(o)$; decode $a$ as 
$(n_1, p)$ --- anchor node and path index
\STATE Partition $\{f_1 \ldots f_F\}$ consecutively from $n_1$; 
\IF{any VNF group cannot be placed} \RETURN \textit{Blocked} \ENDIF
\STATE Decompose path: $s \to n_1 \to [n_2, \ldots] \to d$
\FOR{each segment $(u, w)$}
    \STATE Select modulation $M$ from $D(u,w)$; compute $n = b_{req} 
    \times \eta(M)$
    \STATE Allocate $n$ slots on $(u,w)$ using core group 
    order $\{0,2,4\} \to \{1,3,5\} \to \{6\}$
    \IF{fails} \RETURN \textit{Blocked} \ENDIF
\ENDFOR
\STATE Reserve compute units at all placed nodes; update state
\RETURN \textit{Accepted}
\end{algorithmic}
\end{algorithm}

\subsection{Baseline Approaches}

All baselines use first-fit spectrum allocation across core groups,
differing only in compute node selection.

\subsubsection{Distance-Based Greedy Heuristics}
Compute nodes are ranked by distance score relative to request endpoints
using two strategies: \textbf{Endpoint-Biased} (proximity to source or
destination) and \textbf{Equidistant} (minimizing $|d(s,v) - d(v,d)|$
for balanced path decomposition). The top-5 candidates are attempted in
order, with VNFs assigned to the first feasible node pair and
spectrum allocated via first-fit on resulting segments.

\subsubsection{Greedy Heuristic Baseline (WMSM-F)}
WMSM-F~\cite{ICCCN2026} selects compute waypoints by jointly minimizing
path length and segment imbalance $\delta_{\mathcal{W}}$:
$\min_{\mathcal{W}} \left( \sum_{i=0}^{|\mathcal{W}|} d(w_i, w_{i+1}) + \lambda \cdot \delta_{\mathcal{W}} \right)$, where $\delta_{\mathcal{W}} = \max_i d(w_i, w_{i+1}) - \min_i
d(w_i, w_{i+1})$, with first-fit spectrum allocation and blocking if
no feasible configuration exists.

\subsubsection{Adapted VONE-DRL Baseline}
Inspired by Doherty~\textit{et al.}~\cite{DohertVONE2023}, who apply
masked DRL to VONE over single-core EONs with separate node and link
mapping agents. Both agents are adapted to the 4-VNF chain setting (the node agent 
selects the anchor compute node and the link agent handles per-segment 
RMCSA) and share the same hyperparameters and training budget as 
PPO-Full, ensuring a fair comparison under identical computational 
constraints.

\section{Numerical Evaluation}

\subsection{Simulation Setup}

Simulations use a 24-node USNET topology with 42 bidirectional links 
(30--200~km), 7 cores per fiber, and 120 slots per core (840 slots/link). 
Spectrum follows one guard slot and core group 
search order $\{0,2,4\} \to \{1,3,5\} \to \{6\}$. Modulation is 
distance-adaptive: 16-QAM ($\leq$500~km, 1~slot/Gbps), QPSK 
($\leq$1000~km, 2~slots/Gbps), and BPSK ($\leq$2000~km, 4~slots/Gbps).

VNF hosting is enabled on 22 of 24 nodes, with each compute node having 150 compute units. Requests arrive as a Poisson process with $\lambda \in 
\{5, 10, \ldots, 75\}$, request holding time is exponentially distributed ($\mu = 0.5$), and each request specifies a bandwidth $b_{req} \sim \mathcal{U}(1, 20)$~Gbps 
and $F=4$ VNFs, each requiring $f_i \sim \mathcal{U}(2, 10)$ normalized compute
units. To stress-test performance under realistic traffic asymmetry, 
30\% of requests are hotspot traffic with source drawn uniformly from 
nodes $\{0, 5, 12\}$ and a random destination, while the remaining 70\% have uniformly random source-destination pairs. Hotspot nodes $\{0, 5, 12\}$ were selected as geographically distributed high-degree nodes in the USNET topology, and the 30\% 
hotspot fraction reflects traffic concentration patterns commonly 
observed in metro and backbone networks~\cite{Akella2003}, where a 
small subset of nodes generates a disproportionate share of demand. Performance is evaluated 
over 10,000 requests across 5 independent seeds.

Five strategies are evaluated: (1)~\textit{PPO-Full}: the proposed 
approach jointly selecting the compute node and the routing path, (2)~\textit{Endpoint-Biased}: distance-based heuristic ranking nodes by proximity to the source or 
destination, (3)~\textit{Equidistant}: distance-based heuristic 
ranking nodes by equidistance from source and destination, and 
(4)~\textit{WMSM-F}: the greedy waypoint-based baseline 
from~\cite{ICCCN2026}, and (5)~\textit{VONE-DRL}: the decoupled 
RL baseline adapted from Doherty~\textit{et al.}~\cite{DohertVONE2023}. 
All heuristic baselines apply first-fit spectrum allocation on the resulting 
path segments.

\subsection{Results and Discussion}

\begin{figure}[!t]
\centering
\includegraphics[width=0.88\linewidth]{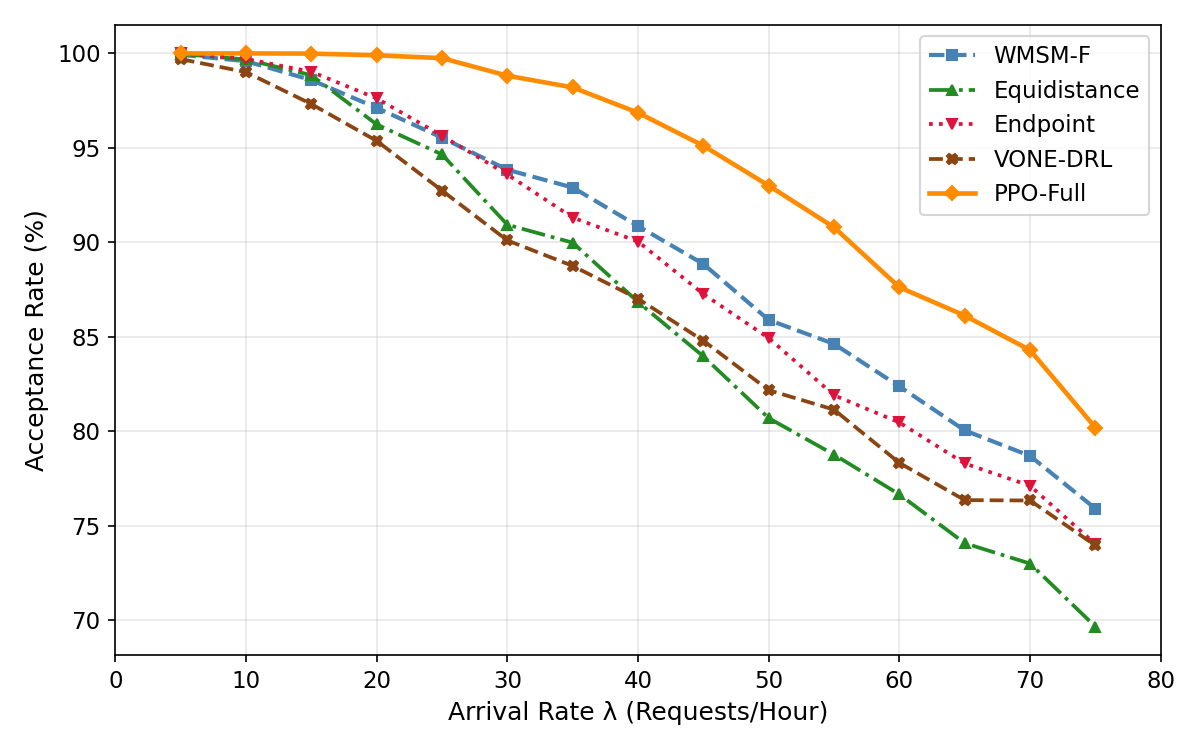}
\vspace{-10pt}
\caption{Acceptance Rate vs. Arrival rates.}
\vspace{-8pt}
\label{fig:acceptance}
\end{figure}

Fig.~\ref{fig:acceptance} shows the acceptance rate under 30\% hotspot 
traffic. PPO-Full maintains the highest acceptance throughout, reaching 
$\approx$80\% at $\lambda=75$, compared to $\approx$76\% for WMSM-F, 
$\approx$74\% for Endpoint-Biased and VONE-DRL, and $\approx$70\% for 
Equidistant. PPO-Full's advantage widens progressively with load, 
confirming that joint node-path optimization adapts more effectively 
to hotspot-induced traffic concentration than deterministic strategies. 
VONE-DRL's low acceptance despite near-zero VONE blocking reveals that 
its decoupled design wastes spectrum resources on inefficient routing.

\begin{figure}[!t]
\centering
\includegraphics[width=0.88\linewidth]{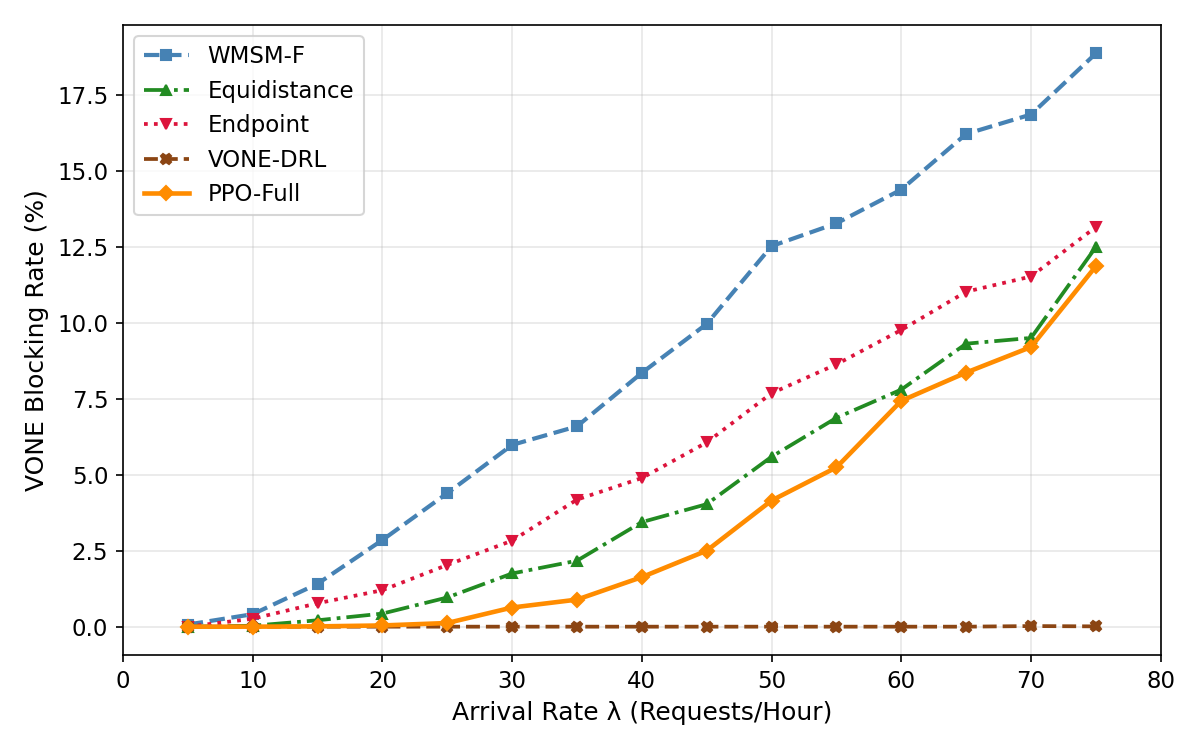}
\vspace{-10pt}
\caption{Blocking Rate for Slice Embedding vs. Arrival rates.}
\vspace{-8pt}
\label{fig:vone_block}
\end{figure}

Fig.~\ref{fig:vone_block} shows VONE blocking. VONE-DRL exhibits 
near-zero VONE blocking as its greedy chain placement always finds a 
feasible compute node, but at the cost of poor spectrum efficiency. 
WMSM-F suffers the highest VONE blocking ($\approx$19\%) as its 
distance-minimizing waypoint selection exhausts compute at 
hotspot-adjacent nodes, while PPO-Full achieves the lowest non-trivial 
VONE blocking ($\approx$12\%), demonstrating that learned node selection 
adapts more effectively to concentrated hotspot demand.

\begin{figure}[!t]
\centering
\includegraphics[width=0.88\linewidth]{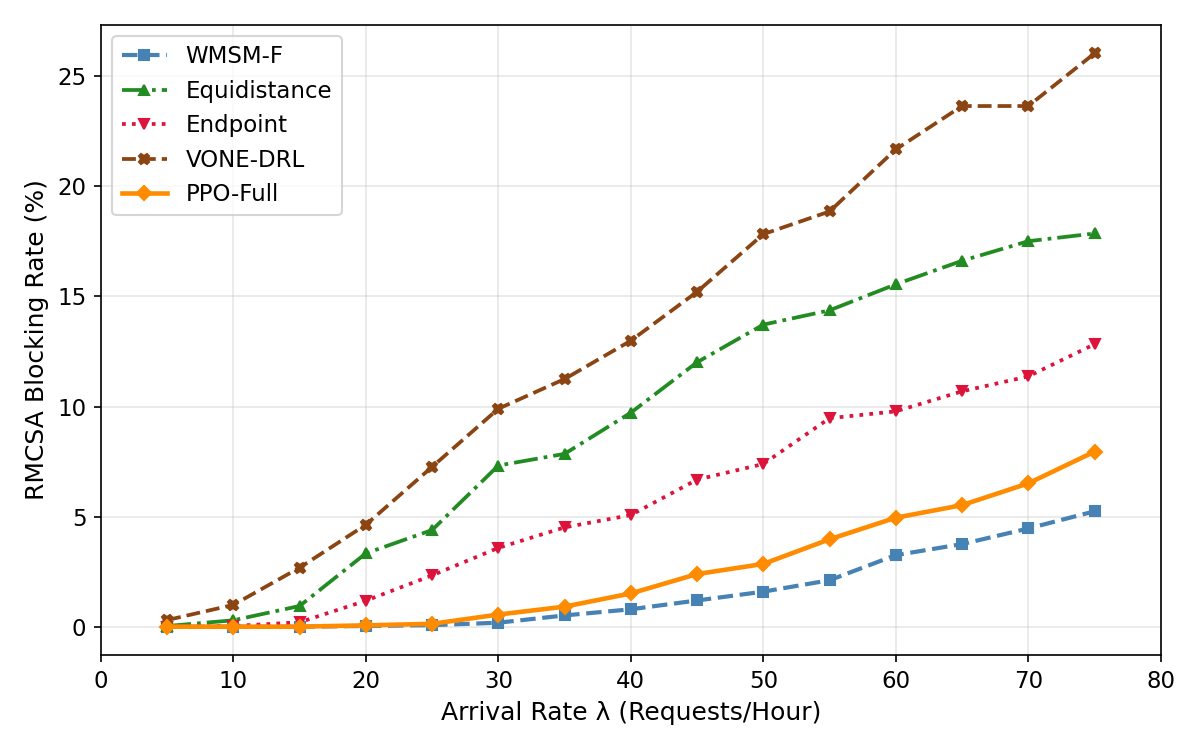}
\vspace{-10pt}
\caption{Blocking Rate for RMCSA vs. Arrival rates.}
\vspace{-8pt}
\label{fig:rmcsa_block}
\end{figure}

Fig.~\ref{fig:rmcsa_block} shows RMCSA blocking. VONE-DRL performs 
worst ($\approx$26\% at $\lambda=75$), confirming its decoupled link 
mapping agent makes spectrum-inefficient decisions. WMSM-F achieves 
the lowest RMCSA blocking ($\approx$5\%) through compact waypoint 
segments, while PPO-Full ($\approx$8\%) outperforms Endpoint-Biased 
($\approx$13\%) and Equidistant ($\approx$18\%).

\begin{figure}[!t]
\centering
\includegraphics[width=0.88\linewidth]{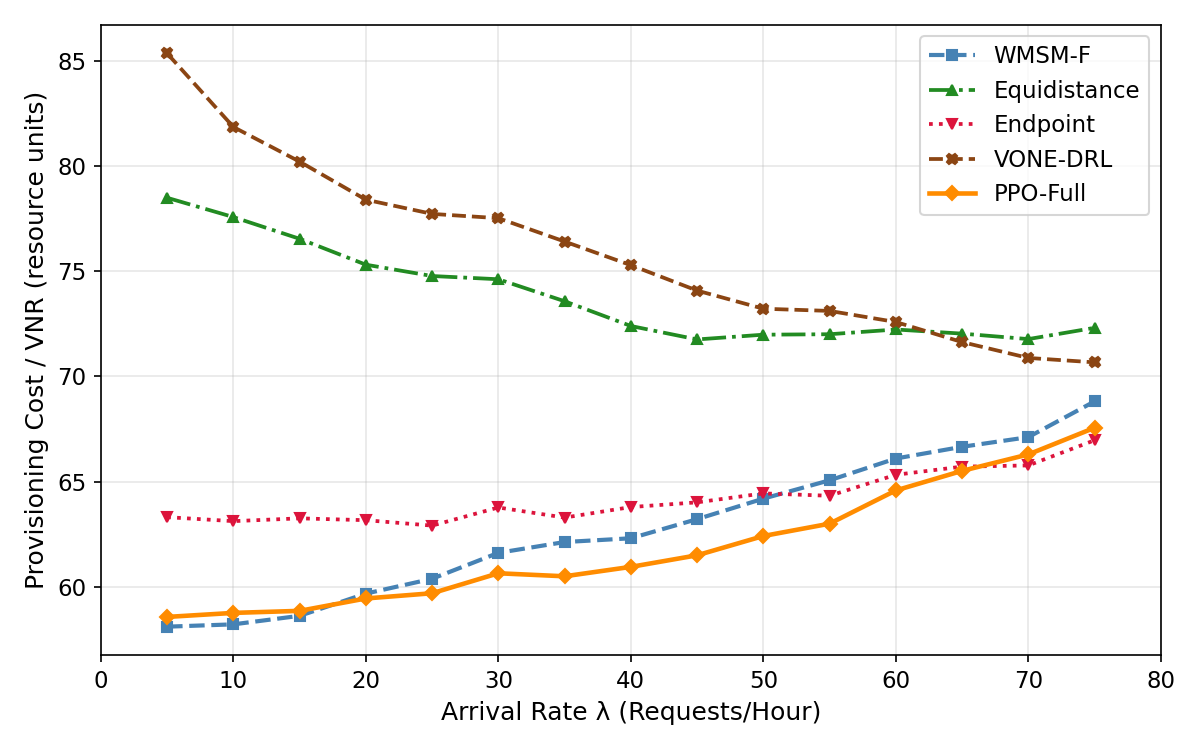}
\vspace{-10pt}
\caption{Average Provisional Cost per VNR vs. Arrival rates.}
\label{fig:cost}
\end{figure}

\begin{figure}[!t]
\centering
\includegraphics[width=0.88\linewidth]{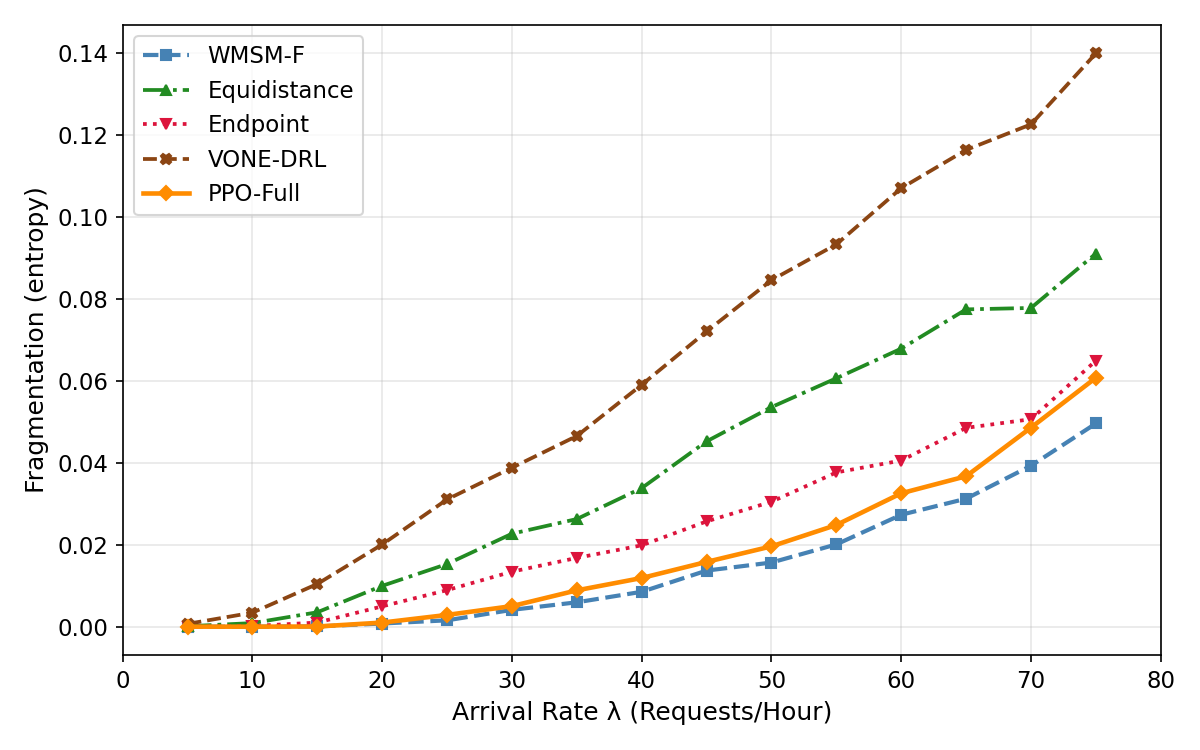}
\vspace{-10pt}
\caption{Fragmentation vs. Arrival rates.}
\label{fig:frag}
\end{figure}

\begin{figure}[!t]
\centering
\includegraphics[width=0.88\linewidth]{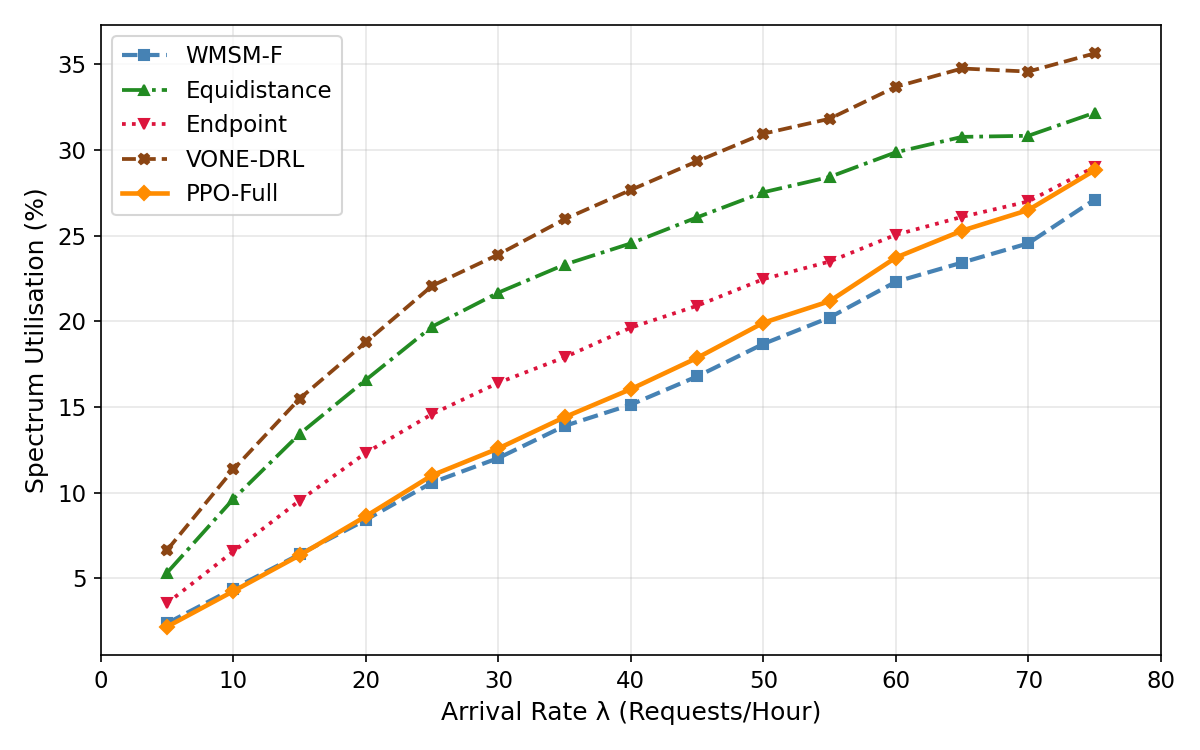}
\vspace{-10pt}
\caption{Spectrum Utilization vs. Arrival rates.}
\vspace{-8pt}
\label{fig:util}
\end{figure}

Fig.~\ref{fig:cost} shows provisioning cost per VNR. VONE-DRL 
incurs the highest cost ($\approx$85 at low load, declining to 
$\approx$71 at $\lambda=75$), while PPO-Full maintains the lowest 
throughout ($\approx$59--67). Equidistant remains consistently high 
($\approx$72--79) due to longer midpoint-biased segments, and WMSM-F 
rises steeply at high load as it is forced onto longer waypoint paths 
under congestion. Fig.~\ref{fig:frag} shows VONE-DRL produces 
the highest fragmentation ($\approx$0.14 at $\lambda=75$), while 
PPO-Full achieves low fragmentation ($\approx$0.06), comparable to 
Endpoint-Biased, and WMSM-F achieves the lowest ($\approx$0.05). 
However, WMSM-F's lower fragmentation is an artefact of its higher 
blocking rate, that is, fewer accepted requests leave larger contiguous free 
spectrum regions, rather than reflecting any genuine spectrum-management 
advantage. Fig.~\ref{fig:util} shows VONE-DRL reaches the 
highest utilization ($\approx$36\%) despite lower acceptance, confirming 
excessive per-VNR spectrum consumption, while PPO-Full maintains 
$\approx$29\%, comparable to WMSM-F. PPO-Full's higher fragmentation 
and utilization relative to WMSM-F reflect a deliberate trade-off: the 
RL agent accepts more requests by routing more aggressively, which 
produces higher fragmentation as a consequence of higher throughput 
rather than inefficient spectrum use.



\section{Conclusion}

This paper proposed PPO-Full, a PPO-based approach for network slice 
embedding in SDM-EONs that jointly optimizes compute node selection 
and RMCSA over $K$ candidate nodes and paths selected via a 
multi-dimensional action space. Evaluated against Endpoint-Biased, 
Equidistant, WMSM-F, and VONE-DRL~\cite{DohertVONE2023} on a 24-node 
USNET topology under hotspot traffic, PPO-Full consistently achieved 
the highest acceptance and lowest per-VNR cost across all load levels. 
WMSM-F's lower fragmentation and RMCSA blocking are artefacts of 
conservative blocking rather than genuine spectrum efficiency, while 
VONE-DRL's decoupled design produced the worst RMCSA blocking despite 
near-zero compute blocking, confirming that joint compute-spectrum 
optimization is essential under realistic heterogeneous traffic. Future 
work will explore spectrum-aware path generation and removal of the 
$K$ restriction through pointer networks or graph attention mechanisms.

\vspace{-6 pt}

\section*{Acknowledgment}
\vspace{-6 pt}
This work was supported in part by the National Science Foundation under Grant No. CNS-2008856.

\bibliographystyle{IEEEtran}
\bibliography{reference}

\end{document}